\documentclass{svproc}
\usepackage{graphicx}
\usepackage{caption}
\usepackage{subcaption}
\usepackage{amsmath}
\usepackage{comment}
\usepackage{float}

\begin{document}
\mainmatter              
\title{Higher-Order Temporal Network Prediction}
\titlerunning{Higher-Order Temporal Network Prediction}  
%
\author{Mathieu Jung-Muller \and Alberto Ceria \and Huijuan Wang}
\authorrunning{Jung-Muller et al.} 
%
\tocauthor{Mathieu Jung-Muller, Alberto Ceria, and Huijuan Wang}
\institute{Delft University of Technology, Mekelweg 4, 2628 CD Delft, Netherlands,\\
\email{H.Wang@tudelft.nl},\\
}

\maketitle              

\begin{abstract}
A social interaction (so-called higher-order event/interaction) can be regarded as the activation of the hyperlink among the corresponding individuals. Social interactions can be, thus, represented as higher-order temporal networks, that record the higher-order events occurring at each time step over time. The prediction of higher-order interactions is usually overlooked in traditional temporal network prediction methods, where a higher-order interaction is regarded as a set of pairwise interactions. We propose a memory-based model that predicts the higher-order temporal network (or events) one step ahead, based on the network observed in the past and a baseline utilizing pair-wise temporal network prediction method. In eight real-world networks, we find that our model consistently outperforms the baseline. Importantly, our model reveals how past interactions of the target hyperlink and different types of hyperlinks that overlap with the target hyperlinks contribute to the prediction of the activation of the target link in the future.

\keywords{higher-order network, temporal network, network prediction, network memory}
\end{abstract}

\section{Introduction}

Temporal networks have been used to represent complex systems with time-varying network topology, where each link between two nodes is activated only when the node pair interacts \cite{holme2012temporal,masuda2016guide}. This classic temporal network presentation assumes interactions to be pairwise. Social contacts/interactions have been mostly studied as pairwise temporal networks. However, contacts/interactions could be beyond pairwise \cite{battiston2020networks}\cite{battiston2021physics}. Individuals may interact in groups \cite{sekara2016fundamental}. A collaboration in scientific paper may engage more than two authors. Such interactions that involve an arbitrary number of nodes are called higher-order interactions/events. Social contacts are thus better represented by higher-order temporal networks.

The classic temporal network prediction problem consists of predicting pairwise contacts one time step ahead based on the temporal network observed in the past. Predicting a temporal network in the future enables better forecast and mitigation of the spread of epidemics or misinformation on the network. The temporal network prediction problem is also equivalent to problems in recommender systems: e.g., predicting which user will purchase which product, or which individuals will become acquaintances at the next time step \cite{lu2012recommender,aleta2020link}. Different methods have been proposed for pairwise temporal network prediction. Some rely on network embeddings: nodes are represented as points in a low dimensional space where connected nodes are supposed to be closer in this embedding space \cite{zhou2019dynamic}. Alternatively, deep learning methods have also been proposed \cite{li2014deep}, for instance, using LSTM methods \cite{chen2019lstm} or adversarial networks \cite{chen2019generative}. However, these methods are at the expense of high computational costs and are limited in providing insights regarding which network mechanisms enable network prediction.
Methods have also been proposed to predict whether a set of nodes will have at least one group interaction in the future \cite{benson2018simplicial,liu2023higher,piaggesi2022effective} and when the first group interaction among these nodes occurs \cite{liu2022neural}.

In this paper, we aim to predict higher-order temporal networks, i.e., higher-order interactions, one time step ahead, based on the higher-order temporal network observed in the past of a duration $L$, and to understand what network properties and which types of previous interactions enable the prediction. Firstly, we explore the memory property of higher-order temporal networks, i.e., to what extent the network topology remains similar over time. Furthermore, we propose a memory-based model to solve the prediction problem utilizing the memory property observed. This model is a generalization to higher-order of the pairwise model proposed in \cite{zou2023memory}. Our model assumes that the activity (interacting or not) of a group at the next time step is influenced by the past activity of this target group and of other groups that form a subset or a superset of the target group. Furthermore, the influence of recent events is considered more impactful than the influence of older events. In the prediction problem,  we assume the total number of events of each group size (order) at the prediction time step is known, and the groups, each of which interact at least once in the network (in the past or future), are known. These assumptions aim to simplify the problem. Beyond, the total number of interactions of each order could be influenced by factors like weather and holiday other than the network observed in the past. The latter assumption means that group friendship is known and we confine ourselves to the prediction of which groups with group friendship interact at the next time step.

We also propose a baseline model that uses a memory-based pairwise temporal network prediction method \cite{zou2023memory}: it considers the higher-order temporal network observed in the past as a pairwise temporal network, predicts pairwise temporal network in the next time step and deduce higher-order interactions from the predicted pairwise interactions at the same prediction time step.

Our model consistently outperforms this baseline in network prediction, as evaluated in eight real-world physical contact datasets. We find that the past activity of the target group is the most important factor for the prediction.
Moreover, the past activity of groups of a large size has, in general, a lower impact on the prediction of events of the target group than groups of a small size.

The rest of the paper is organized as follows. We introduce in Section \ref{section-definitions} the representation of higher-order temporal networks and in Section \ref{section-datasets} the datasets we use to design and evaluate our prediction method. Key temporal network properties are explored in Section \ref{section-jaccard} to motivate the design of our model, which is explained in Section \ref{section-models}. In Section \ref{section-eval}, our model is evaluated and compared to our baseline and further interpreted. 

\section{Network representation} \label{section-definitions}

A pairwise temporal network \(G\) measured at discrete times can be represented as a sequence of network snapshots \( G= \{G_1, G_2, ..., G_T \}  \), where \(T\) is the duration of the observation window, \( G_t = (V, E_t) \) is the snapshot at time step \(t\) with \(V\) and \(E_t\) being the set of nodes and interactions, respectively. If nodes \(a\) and \(b\) have a contact at time step \(t\), then \( (a, b) \in E_t\). Here, we assume all snapshots share the same set of nodes \(V\). The set of links in the time-aggregated network is defined as \( E = \bigcup_{t=1}^{t=T} E_t \). A pair of nodes is connected with a link in the aggregated network if at least one contact occurs between them in the temporal network. 
The temporal connection or activity of link \(i\) over time can be represented by a \(T\)-dimension vector \(x_i\) whose element is \(x_i(t)\), where \( t \in [1, T]\), such that \(x_i(t) = 1\) when link \(i\) has a contact at time step \(t\) and \(x_i(t) = 0\) if no contact occurs at \(t\). A temporal network can thus be equivalently represented by its aggregated network, where each link \(i\) is further associated with its activity time series \(x_i\).

Social interactions, which may involve more than two individuals, can be more accurately represented as a higher-order temporal network \(H\), which is a sequence of network snapshots \(H = \{H_1, ..., H_T \}\), where \( H_t = (V, {\mathcal{E}}_t) \) is the snapshot at time step \(t\) with \(V\) being the set of nodes shared by all snapshots and \( {\mathcal{E}_t}\) the set of hyperlinks that are activated at time step \(t\).
The activation of a hyperlink \( (u_1, ..., u_d) \) at time step \(t\) corresponds to a group interaction among nodes \(u_1\), ..., \(u_d\) at time step \(t\). The hyperlink \( (u_1, ..., u_d) \) active at time step \(t\) is called an event or interaction and its size is \(d\).
If $h_1 \subset h_2$, i.e., \(h_1\) is included in \(h_2\), we call \( h_2\) a super-hyperlink of \(h_1\) and \(h_1\) a sub-hyperlink of \(h_2\).
The set of hyperlinks in the higher-order time-aggregated network is defined as \( \mathcal{E} = \bigcup_{t=1}^{t=T} {\mathcal{E}}_t \). A hyperlink belongs to \(\mathcal{E}\) if it is activated at least once in the temporal network.
 A higher-order temporal network can thus be equivalently represented by its higher-order aggregated network, where each hyperlink \(i\) is further associated with its activity time series \(x_i\).

\section{Datasets} \label{section-datasets}
\begin{table}[h!]
    \centering
    \begin{tabular}{|c|c|c|c|c|c|}
        \hline
        Dataset & Order 2 & Order 3 & Order 4 & Order 5 & Order 6+ \\
        \hline\hline
        Science Gallery & 12770 & 1421 & 77 & 7 & 0 \\ \hline
        Hospital & 25487 & 2265 & 81 & 2 & 0 \\ \hline
        Highschool 2012 & 40671 & 1339 & 91 & 4 & 0 \\  \hline
        Highschool 2013 & 163973 & 7475 & 576 & 7 & 0 \\ \hline
        Primaryschool & 97132 & 9262 & 471 & 12 & 0 \\ \hline
        Workplace & 71529 & 2277 & 14 & 0 & 0 \\ \hline
        Hypertext 2009 & 18120 & 874 & 31 & 12 & 4 \\ \hline
        SFHH Conference & 48175 & 5057 & 617 & 457 & 199 \\ \hline 
    \end{tabular}
\caption{ Number of events of every order for every dataset after preprocessing. }
\label{tab:dataset_table}
\end{table}

To design and evaluate our temporal network prediction methods, we consider eight empirical physical contact networks from the SocioPatterns project\footnote{http://www.sociopatterns.org/}. They are collections of face-to-face interactions at a distance smaller than 2 m, in several social contexts such as study places (Highschool2012 \cite{fournet2014contact}, Highschool2013 \cite{mastrandrea2015contact}, Primaryschool \cite{stehle2011high}), conferences (SFHH Conference \cite{cattuto2010dynamics}\cite{stehle2011simulation}, Hypertext2009 \cite{isella2011s}), workplaces (Hospital \cite{vanhems2013estimating}, Workplace \cite{genois2018can}) or an art gallery (Science Gallery \cite{isella2011s}). These face-to-face interactions are recorded as a set of pairwise interactions. Based on them, group interactions are deduced by promoting every fully-connected clique of \(\binom{d}{2}\) contacts happening at the same time step to an event of size \(d\) occurring at this time step. Since a clique of order \(d\) contains all its sub-cliques of order \(d' < d \), only the maximal clique is promoted to a higher-order event, whereas sub-cliques are ignored. This method has been used in \cite{cencetti2021temporal} and \cite{ceria2023temporal}. The datasets are also preprocessed by removing nodes not connected to the largest connected component in the pairwise time-aggregated network and long periods of inactivity, when no event occurs in the network. Such periods usually correspond, e.g., to night and weekends, and are recognized as outliers. This corresponds to the preprocessing done in \cite{ceria2022topological} and \cite{ceria2023temporal}. The total number of events of each order in our datasets, after preprocessing, is shown in Table \ref{tab:dataset_table}.

\section{Network memory property} \label{section-jaccard}

\begin{figure*}[h!]
    \centering
    \subcaptionbox{Order 2\label{fig:jaccard_2}}{%
        \includegraphics[width=0.48\textwidth]{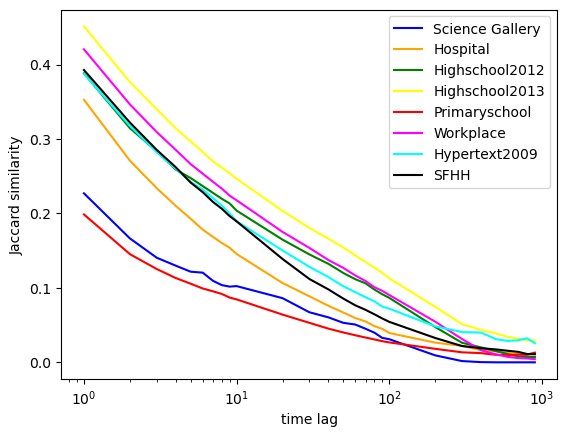}
    }
    \subcaptionbox{Order 3\label{fig:jaccard_3}}{%
        \includegraphics[width=0.48\textwidth]{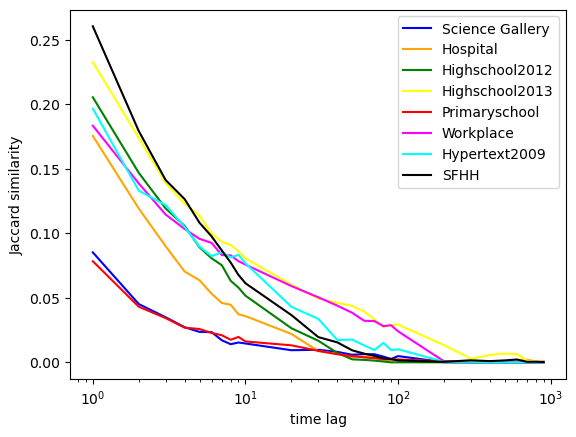}
    }
    \subcaptionbox{Order 4\label{fig:jaccard_4}}{%
        \includegraphics[width=0.48\textwidth]{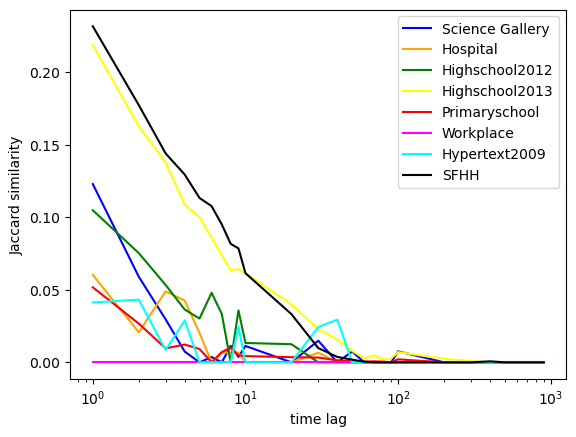}
    }
    \subcaptionbox{Order 5\label{fig:jaccard_5}}{%
        \includegraphics[width=0.48\textwidth]{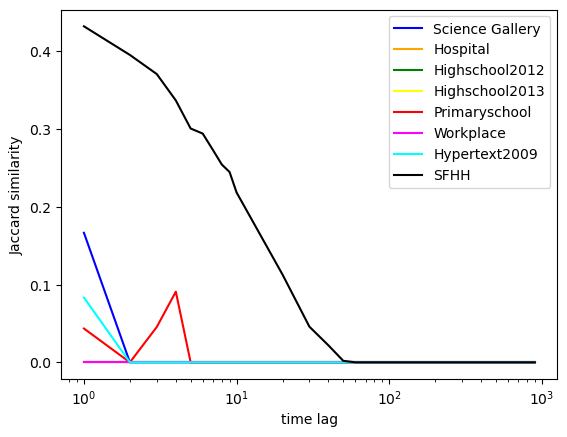}
    }
    \caption{Jaccard similarities of a network at two time steps for each order as a function of the time lag in eight real-world physical contact networks.}
    \label{fig:jaccard}
\end{figure*}

Zou et al. observed properties of time-decaying memory in pairwise temporal networks. This means that different snapshots of the network share certain similarities. These properties were used to better predict pairwise interactions \cite{zou2023memory}. Inspired by this, we also want to know whether higher-order temporal networks have memory at different orders and whether this property can be used to predict higher-order events in temporal networks.

Therefore, we examine the Jaccard similarity of the network at two different time steps and for every order. The Jaccard similarity measures how similar two given sets are by taking the ratio of the size of the intersection set over the size of the union set. In our case, we compute the Jaccard similarity \( \frac{ |\mathcal{E}_{t_1}^n \cap \mathcal{E}_{t_2}^n| }{ |\mathcal{E}_{t_1}^n \cup \mathcal{E}_{t_2}^n| } \) for every order \(n\), between the set \( \mathcal{E}_{t_1}^n \) of \(n\)-hyperlinks (hyperlinks of order \(n\)) active at a time step \(t_1\) and the set \( \mathcal{E}_{t_2}^n \) of \(n\)-hyperlinks active at a time step \(t_2\), called \( \mathcal{E}_{t_2}^n \). The difference \(t_2-t_1\) is called the time lag.

As shown in Figure \ref{fig:jaccard}, the similarity decays as the time lag increases for orders 2, 3, and 4, respectively, in all datasets. The time-decaying memory at order 5 has been observed only in SFHH,  the only network that has a non-negligible number of order 5 events, as shown in Table \ref{tab:dataset_table}.

\section{Models} \label{section-models}

\subsection{Baseline}

We propose a baseline for higher-order temporal network prediction utilizing the following pairwise network prediction model, called Self-Driven SD model, proposed in \cite{zou2023memory}. The SD model is a memory-based model that predicts a pairwise link's activity at the next time step based on its past activity. The SD model estimates the tendency \( w_j(t+1) \) for each link \(j\) to be active at time \(t+1\) as:
\newline\newline
\( w_j(t+1)= \sum^{k=t}_{k=t-L+1} x_j(k) \exp^{-\tau(t-k)} \),
\newline\newline
where \(t+1\) is the prediction time step, \(L\) is the length of the past observation used for the prediction, \(\tau\) is the exponential decay factor, and \(x_j(k)\) is the activation state of link \(j\) at time \(k\): \( x_j(k)=1 \) if the link \(j\) is active at time \(k\) and \( x_j(k)=0 \) otherwise.

The activation tendency is computed for each link in the pairwise aggregated network, which is given. Given the number of pairwise contacts \(n_{t+1}^{2}\) occurring at \(t+1\), the \(n_{t+1}^{2}\)  links with the highest activation tendency at \(t+1\) will be predicted to have contacts at \(t+1\).

We propose a baseline model that firstly considers the higher-order temporal network observed in the past of duration L as pairwise temporal network, then applies the pairwise memory-based model to predict pairwise interactions at the prediction step, and deduces higher-order interactions from the predicted pairwise interactions, using the same method that promotes interactions that form a clique to a higher-order event (see Section \ref{section-datasets}). This set of higher-order interactions is considered the prediction made by the baseline model.

\subsection{Generalized model}

The time-decaying memory is also observed at different orders in the higher-order temporal networks. This motivate us to generalize the SD model for higher-order network prediction. The essence of the generalized model is that the future activity of a hyperlink should be dependent on its past activity. Furthermore, it has been shown that events of different orders that occur close in time tend to overlap in component nodes \cite{ceria2023temporal}. Hence, the activity of a hyperlink is also supposed to be dependent on the past activity of its sub-hyperlinks and super-hyperlinks. Finally, recent events should have more influence than older events, based on the observed time-decaying memory.

Therefore, we propose the activation tendency of a hyperlink \(j\) at a prediction time step\(t+1\) as:
\newline\newline
\( w_j(t+1)= \sum^{k=t}_{k=t-L+1} \sum_{i \in S_j} c_{d_i d_j} x_i(k) \exp^{-\tau(t-k)} \),
\newline\newline
where \(L\) is the length in time of the network observed in the past used for the prediction, \(\tau\) is the exponential decay factor, \(S_j\) is the set of hyperlinks that are either sub-hyperlinks or super-hyperlinks of \(j\), and \(x_i(k)\) is the activation state of hyperlink \(i\) at time \(k\). \( c_{d_i d_j} \) is the coefficient of cross-order influence, for which \( d_i \) is the size of hyperlink \(i\) and \( d_j \) is the size of hyperlink \(j\). For instance, \( c_{32} \) is the coefficient associated with the influence of the activation of a 3-hyperlink on the activation of one of its sub-hyperlinks of size 2. We put \(c_{dd}=1\) for any arbitrary hyperlink size \(d\). Different sub-models of our general model are obtained by varying the values of the cross-order coefficients \( c_{d_1d_2} \) for \( d_1 \neq d_2 \).

 The activation tendency is computed for each hyperlink in the higher-order aggregated network, which is given in the prediction problem. Given the number \(n_{t+1}^{o}\) of events of each order \(o\) at the prediction step $t+1$, the \(n_{t+1}^{o}\) hyperlinks of order \(o\) with the highest activation tendency at \(t+1\) are predicted to be active.


\section{Model evaluation} \label{section-eval}

\subsection{Network prediction quality}

We aim to predict the higher-order temporal network at time step \(t+1\) based on the network observed between \(t-L+1\) and \(t\). To evaluate the proposed prediction methods, every time step within $[L+1,T]$ is chosen as a possible prediction step, i.e., $t+1 \in [L+1,T]$, where \(T\) is the global lifespan of each empirical network. The prediction quality of a given model for events of an arbitrary order is evaluated via prediction accuracy: the ratio between the total number of true positives (correctly predicted events) over all possible prediction steps $[L+1,T]$ for that order and the total number of events of that order that occur within $[L+1,T]$.

\subsection{Parameter choice of the generalized model} \label{model-parameters}

Since the events of orders higher than 4 are few in number in real-world physical contact networks considered, we focus on the prediction of events of orders 2, 3, and 4, respectively based on the previous activities of events of orders 2, 3, and 4. For every order (e.g., order 3), we make its associated pair of cross-order coefficients (e.g.,\(c_{23}\), \(c_{43}\)) take all possible values in \( \{ 0.0, 0.1, ..., 1.0 \} \times \{ 0.0, 0.1, ..., 1.0 \} \). Cross-order coefficients larger than 1 or smaller than 0 lead to lower prediction quality in general and are, therefore, not considered.

We choose the duration of the past network observation \(L=30\) for prediction, which is equivalent to 600s in our real-world physical contact networks (the duration of each time step is 20ms). This choice was found by comparing the accuracy of the prediction for different values of \(L\) between 1 and \(T/2\) and for different values of \( \tau \) between 0.25 and 1. Because a small L means low computational complexity, we identify L=30 as the smallest L that does not lead to an evident drop in prediction quality compared with \(L=T/2\).


We compared the prediction performance for different values of the decay factor \(\tau\), where \( \tau \in \{0, 0.25, ..., 1.5\} \). When \(\tau=0\), i.e., previous interactions contribute equally in predicting the connection tendency of a hyperlink independent of when these interactions occur, the model performs the worst in every order for all datasets. This is in line with the time-decaying memory we have observed. In general, the choice of $\tau \in [0.25, 1.5]$ has little influence in the performance. We will focus on the performance analysis of the generalized model in comparison with the baseline model when \(\tau=0.25\) in this paper, since other choices of $\tau$ lead to the same observation.

\subsection{Performance Analysis} \label{prediction-performance}

For every order 2, 3, or 4, we compute the prediction accuracy obtained by our generalized model with any pair of cross-order coefficients in \( \{ 0.0, 0.1, ..., 1.0 \} \times \{ 0.0, 0.1, ..., 1.0 \} \) and identify the best performance achieved by choosing the optimal coefficient pair. This best performance, referred as the prediction accuracy of the model is further compared with the prediction quality of the baseline. As shown in Table \ref{tab:best-results}, our generalized model performs overall better than the baseline for the interaction prediction for any order and in each network. The outperformance is more evident at orders 3 and 4.

\begin{table}[t]
    \centering
    \begin{tabular}{|c|c|c|c|}
        \hline
        Dataset & Order 2 & Order 3 & Order 4\\
        \hline\hline
        Science Gallery & 0.33 (0.33) & \textbf{0.23} (0.16) & \textbf{0.57} (0.22) \\ \hline
        Hospital & \textbf{0.53} (0.52) & \textbf{0.50} (0.32) & \textbf{0.72} (0.17) \\ \hline
        Highschool2012 & \textbf{0.56} (0.55) & \textbf{0.50} (0.38) & \textbf{0.67} (0.19) \\  \hline
        Highschool2013 & 0.61 (0.61) & \textbf{0.40} (0.34) & \textbf{0.61} (0.36) \\ \hline
        Primaryschool & 0.32 (0.32) & \textbf{0.19} (0.16) & \textbf{0.33} (0.09) \\ \hline
        Workplace & \textbf{0.57} (0.56) & \textbf{0.49} (0.30) & \textbf{0.50} (0.07) \\ \hline
        Hypertext2009 & 0.50 (0.50) & \textbf{0.52} (0.34) & \textbf{0.68} (0.10) \\ \hline
        SFHH Conference & \textbf{0.53} (0.52) & \textbf{0.45} (0.38) & \textbf{0.58} (0.39) \\ \hline
    \end{tabular}
\caption{ Prediction accuracy of the generalized model and the baseline (in parentheses) per order for every dataset. The prediction accuracy of the generalized model is in bold if it is larger than that of the baseline model .}
\label{tab:best-results}
\end{table}

\begin{figure*}[ht!]
    \centering
    \includegraphics[width=0.48\textwidth]{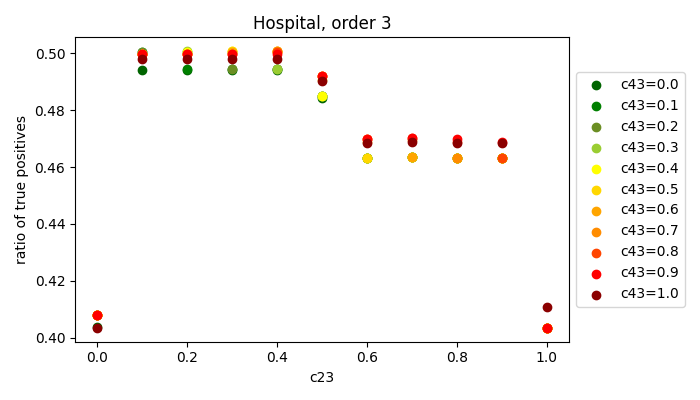}
    \includegraphics[width=0.48\textwidth]{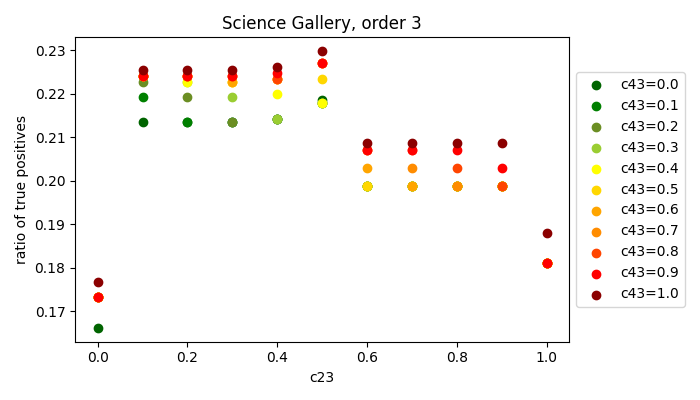}
    \includegraphics[width=0.48\textwidth]{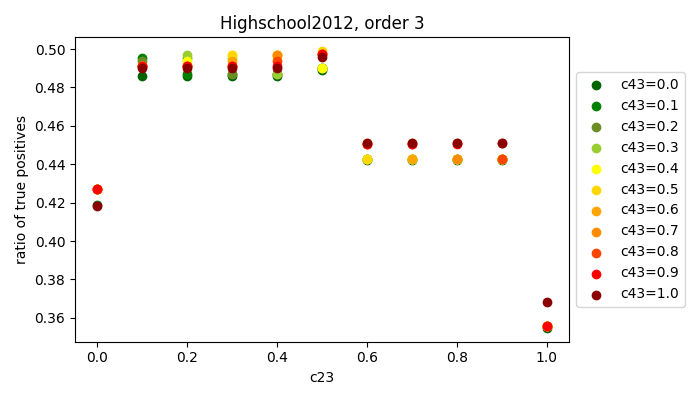}
    \includegraphics[width=0.48\textwidth]{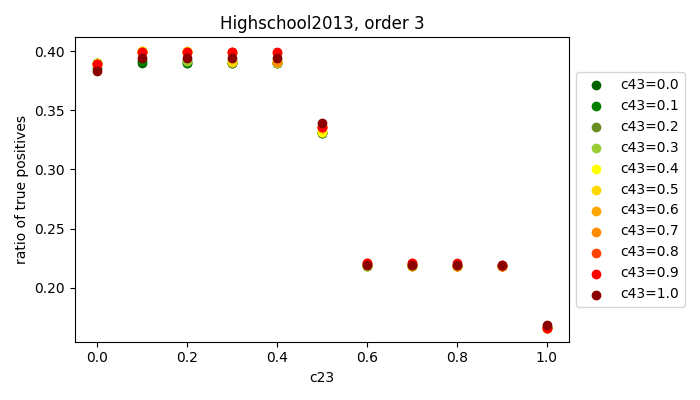}
    \includegraphics[width=0.48\textwidth]{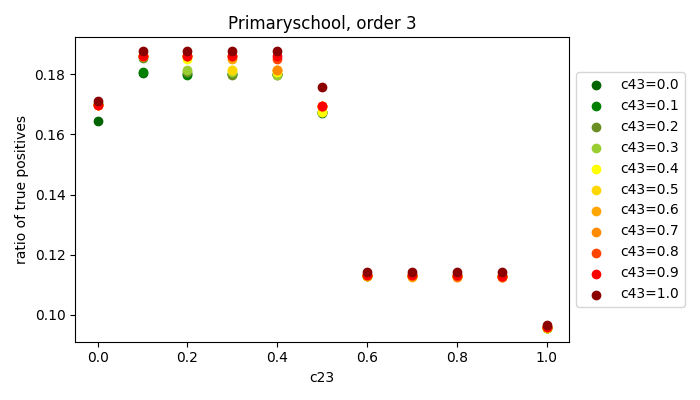}
    \includegraphics[width=0.48\textwidth]{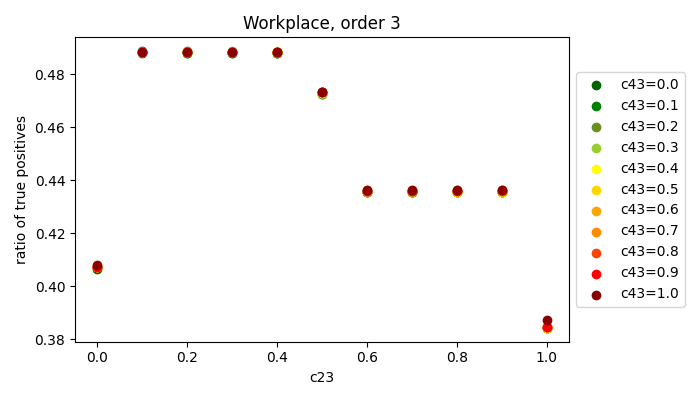}
    \includegraphics[width=0.48\textwidth]{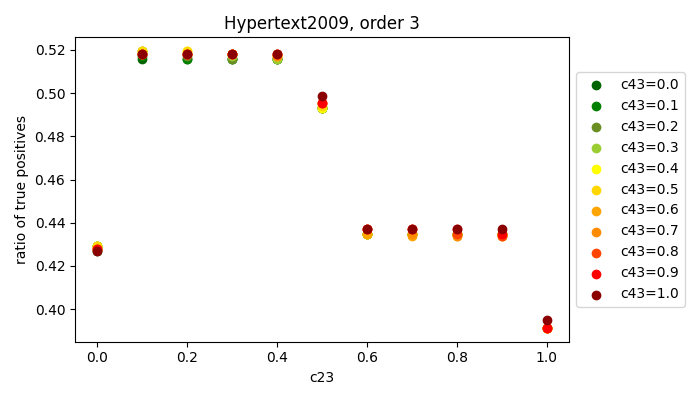}
    \includegraphics[width=0.48\textwidth]{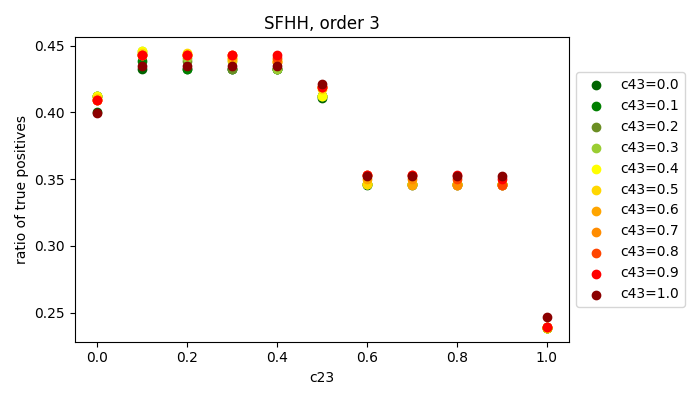}
    \caption{ Precision (Ratio of true positives) of predicting events of order 3 as a function of \(c_{23}\), with \(c_{43}\) fixed, for all datasets. }
    \label{fig:swoc-23-appendix}
\end{figure*}

\begin{figure*}[ht!]
    \centering
    \includegraphics[width=0.48\textwidth]{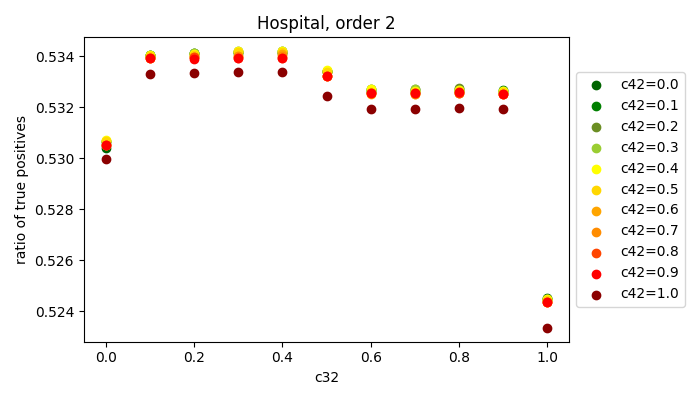}
    \includegraphics[width=0.48\textwidth]{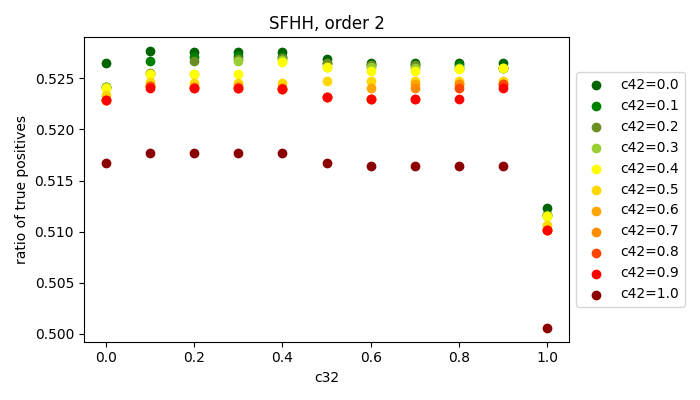}
    \includegraphics[width=0.48\textwidth]{images/score_wrt_one_coeff/Hospital_3_c23.png}
    \includegraphics[width=0.48\textwidth]{images/score_wrt_one_coeff/SFHH_3_c23.png}
    \includegraphics[width=0.48\textwidth]{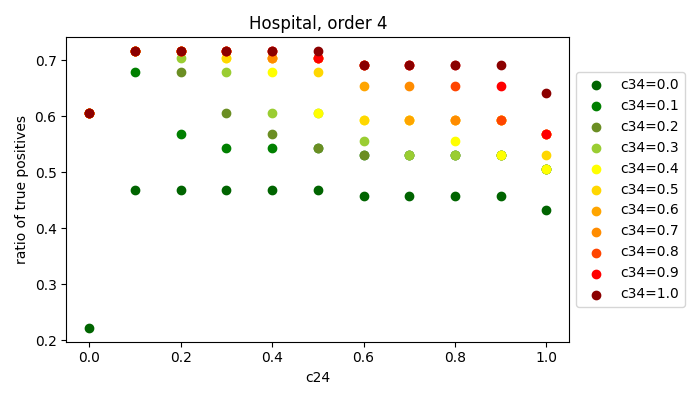}
    \includegraphics[width=0.48\textwidth]{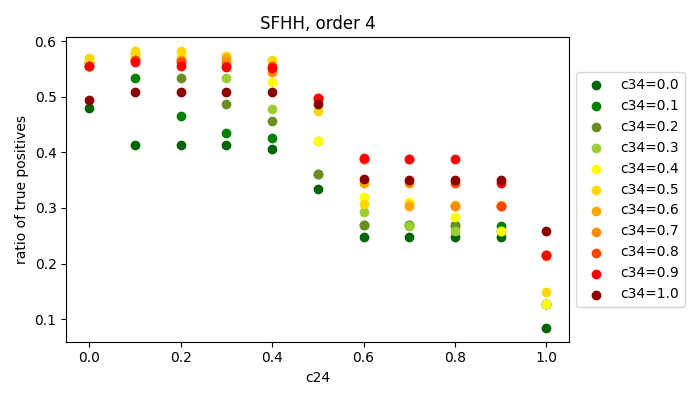}
    \caption{Precision (Ratio of true positives) of predicting events of a given order as a function of one coefficient, with the other coefficient fixed, for two datasets. }
    \label{fig:swoc}
\end{figure*}

For the prediction of events of a given order, the two corresponding cross-order coefficients of the generalized model affect the prediction accuracy, as shown in Figure \ref{fig:swoc-23-appendix} and \ref{fig:swoc}. We will explore which coefficient ranges tend to lead to optimal prediction accuracy. This will enable us to understand how events of different types of hyperlinks (super- and sub-hyperlinks) in the past contribute to the prediction of the event of a target hyperlink.

Take the prediction of events of order 3 as an example. Figures \ref{fig:swoc-23-appendix} and \ref{fig:swoc} show that close to optimal prediction precision is obtained approximately when $c_{23} \in \{ 0.1, ..., 0.4 \}<c_{33}=1$. This means the interaction of a sub-hyperlink of order 2 in the past leads to a lower activation tendency of the target hyperlink of order 3 compared to the interaction of the target hyperlink occurring at the same time in the past. The influence of $c_{43}$ on the prediction quality is not evident, likely due to the small number of events of order 4. For events of order 2, the prediction quality tends to be optimal when \( c_{32}, c_{42} \in \{ 0.1, ..., 0.4 \} \), though their influence on prediction quality is less evident, as partially shown in Figure \ref{fig:swoc}. For order 4 (see also Figure \ref{fig:swoc}), $c_{24}$ and $c_{34}$ affect the prediction quality evidently and $c_{24} \in \{ 0.1, ..., 0.4 \}$ and $c_{34} \in \{0.6, ..., 0.9\}$ tend to give rise to close to optimal prediction precision. To achieve the optimal prediction quality, $c_{24}<c_{34}<c_{44}=1$, and $c_{23}<c_{33}$. This means that the activation of a hyperlink that overlaps more with the target hyperlink in nodes implies a higher activation tendency of the target link in the future. In general, the activation of super- and sub-hyperlinks all contribute to the activation of the target hyperlink in the future, since cross-order coefficients zero lead to worse prediction precision. However, the choice of the contribution of a super-hyperlink activation (i.e., $c_{d_i d_j}$ when $d_i>d_j$) in predicting the activity of target hyperlink $j$ does not affect the prediction quality evidently, likely because of the relatively small number of activations of a super-hyperlink compared to the number of activations of the target link.


\section{Conclusion and Discussion}

In this paper, we proposed a network-based higher-order temporal network prediction model that predicts the activity of each higher-order hyperlink at the next time step based on the past activity of this hyperlink and of its sub- and super-hyperlinks. The contributions of the different hyperlinks are weighted with an exponential decay depending on how far in the past the events occurred. Our model was shown to perform consistently better than the baseline directly derived from a pairwise prediction method. 
Using the previous activities of the target link itself, its sub- and super-hyperlinks together enable better prediction precision than considering only the activity of the target link itself, which is though the most influential factor. A past event of a sub-hyperlink that overlap more with the target hyperlink in nodes suggests a higher activation tendency of the target link. The contributions of super-hyperlinks (when the corresponding cross-order coefficients vary with the range $[0,1]$) does not affect the prediction quality evidently, likely because of the relatively small number of events of super-hyperlinks in comparison with that of the target hyperlink.

We have focused on higher-order social contact networks that are derived from measurement of pairwise interactions. These higher-order networks have the property that a hyperlink and its sub-hyperlink are never activated at the same time, a property shared by the higher-order network predicted by the baseline model. It would be interesting to explore the proposed methods and better methods to solve the network prediction problem in other types of higher-order temporal networks, that may not have this property nor the memory property. The baseline model could be improved by, e.g., using the total number of events of each order in the prediction step, which has already utilized by our generalized model.

%
%
\section*{Acknowlegement}
This publication is part of the project FORT-PORT (with project number KICH1.VE03.21.008 of the research programme KIC - MISSION 2021 which is (partly) financed by the Dutch Research Council (NWO).
\bibliographystyle{splncs03}
\bibliography{bibliography}

\end{document}